\documentclass[conference]{IEEEtran}

\usepackage{listings}
\usepackage{bytefield}
\usepackage{parskip}
\usepackage{float}
\usepackage{graphicx}
\graphicspath{ {./Images/} }
\usepackage{multirow}
\usepackage{tabularray}

\begin{document}

\title{EJ-FAT Joint ESnet JLab FPGA Accelerated Transport Load Balancer}

\author{
    Stacey Sheldon \textit{(ESnet) stac@es.net}
    Yatish Kumar   \textit{(ESnet) yak@es.net}   \\
    Michael Goodrich \textit{(Jefferson Lab) goodrich@jlab.org}
    Graham Heyes \textit{(Jefferson Lab) heyes@jlab.org}
}

\maketitle

\begin{abstract}
To increase the science rate for high data
rates/volumes, Thomas Jefferson National Accelerator Facility
(JLab) has partnered with Energy Sciences Network (ESnet) to
define an edge to data center traffic shaping / steering transport
capability featuring data event-aware network shaping and
forwarding.

The keystone of this ESnet JLab FPGA Accelerated
Transport (EJFAT) is the joint development of a
dynamic compute work Load Balancer (LB) of UDP streamed
data. The LB is a suite consisting of a Field Programmable Gate
Array (FPGA) executing the dynamically configurable, low fixed
latency LB data plane featuring real-time packet redirection at
high throughput, and a control plane running on the FPGA host
computer that monitors network and compute farm telemetry in
order to make dynamic decisions for destination
compute host redirection / load balancing. 

The LB provides for three forms of scaling. It provides horizontal scale by adding more FPGAs for increased bandwidth. Second it scales out to the number of core compute hosts independent of the number of source DAQs. Thirdly it allows for a flexible number of CPUs and threads per host, treating each receiving thread as an independent LB destination. The LB provides seamless integration of edge / core computing to support direct experimental data processing. Immediate use will be at JLab science programs and others such as the EIC (Electron Ion Collider). Data centers of the future will need high throughput and low latency for both live streamed and recorded data for running experiment data acquisition analysis and data center use cases.

EJ-FAT is a development for production use within DOE.  When completed, it will have an operational impact for integrated research infrastructure as called for in \cite{b8}, \cite{b9}, and \cite{b10}.  It demonstrates a new load balancing architecture, when compared with prior solutions like Server Load Balancing.

\end{abstract}


\section{Introduction}
\IEEEPARstart{T}{he} US Department of Energy (DOE) operates a number of large science facilities, such as particle accelerators, x-ray light sources and electron microscopes.  Each facility is instrumented with many high speed A/D data acquisition systems (DAQs) that can produce multiple 100 Gbps data streams for recording and processing.  This processing is conducted on large banks of compute nodes (CN) in local and remote data facilities.  This paper describes the design of a real time load balancer for distributing UDP encapsulated DAQ payloads into a dynamically allocated set of compute elements.

The load balancer protocol is designed to support Wide Area Network (WAN) latencies for geographically distributed accelerator facilities and high performance computing centres.

\subsection{Load Balancing System Description}

\begin{figure}
    \centering
    \includegraphics[width=\columnwidth]{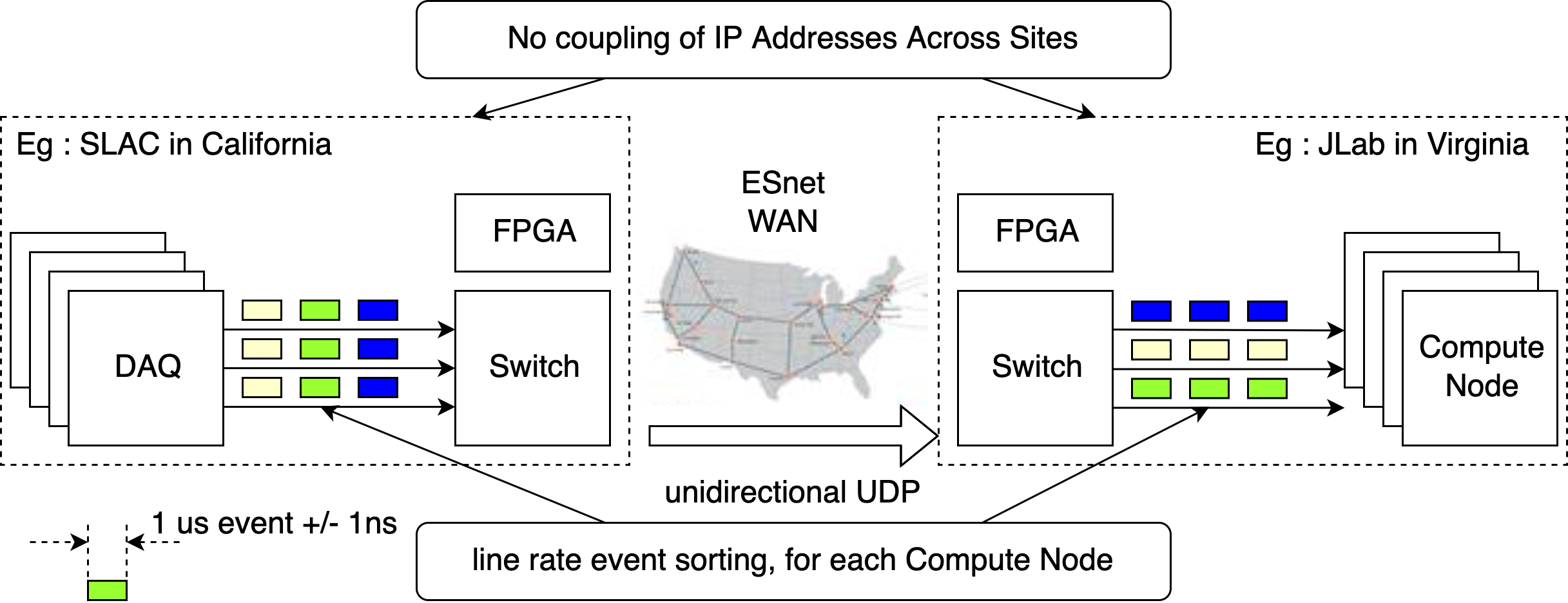}
    \caption{DAQ to Compute Node Dataflow}
    \label{fig:EndtoEndFlow}
\end{figure}

Fig. \ref{fig:EndtoEndFlow} describes the end to end data flow from the DAQ's to the compute nodes in two different facilities separated by a wide area network.  Usually a smaller number of DAQs are processed by a larger number of compute nodes.  As well the number and location of the compute nodes can change when the experiment is run, and in some cases compute nodes can be added or removed while the experiment is running.  

A compute facility can host the processing for a number of different accelerators or experiments.  Each has a unique format for its event data, that is often driven by the physical architecture of the accelerator and its scientific purpose.  3D particle tracking is very different from 2D imaging for example. 2D data is usually grouped in 2D image frames, whereas 3D particle tracking is often captured as a linear time sequence, from which path tracking algorithms can construct a 3D path. The load balance transport needs to be agnostic to these differences, whilst still delivering related pieces of the payloads to the correct computational element for a coherent reassembly process.  As well a compute facility might host more than one experiment,  making it necessary for the load balancer to support multiple sets of flows, without interference or leakage between experiments.

In order to achieve these goals we established EJ-FAT's design objectives as described in the next section.

\subsection{Load Balancing Design Objectives}
The load balancer design is guided by several operational objectives that are driven by the need to source traffic from many different national laboratories to several compute facilities, across a wide area network. \\


\subsubsection{Separation of IP Addresses}
It has been common practice to program the IP address of the compute nodes into the DAQ systems that source IP packets. The DAQ can easily build up a UDP packet that can be routed directly to the compute node.  This has the disadvantage that two different laboratories separated by distance, and site boundaries need to be closely coupled.  As well it is brittle in terms of propagating changes in the compute infrastructure in one lab, to the the sources in a different lab.  Finally it is restrictive, in terms of the mapping from N DAQs to M compute nodes.  In 3D particle tracking for example, every DAQ needs to spread traffic on every CN.  So each DAQ needs to synchronize and maintain state for which current DAQ sample is destined for which CN.  

By introducing a load balancer in the middle, we completely decouple the IP addresses of the compute nodes from the DAQs. The DAQs only need to be provisioned with one IP address, which is the address of the load balancer itself.  Packets leaving the load balancer have their source IP set to the load balancer node and destination IP address rewritten to the addresses of the compute nodes.  The CN addresses are known only to the load balancer and are downloaded at the beginning of an experiment.

\subsubsection{In network sorting of Event Data}
In a triggered DAQ system, it is common for multiple DAQ channels to observe a single event or particle at the same time.  All of those observations at a particular trigger time, should be directed to the same compute node.  This allows the event analysis to happen, without the need for inter node communication.  The load balancer achieves this by making balancing decisions taking into account an experiment specific marker that identifies related pieces of data.

\subsubsection{Stateless load balancing}
In order to provide simplicity, speed and horizontal scale, we established a goal that the load balancer would be stateless.  The LB protocol is designed so that the load balancer can examine a single packet, with no other history, and determine its final destination.

\subsubsection{Compute node feedback for dynamic LB}
Once an experiment starts running, for various reasons some compute nodes will be faster or slower than others.  The load balancer needs a mechanism to change the weighting of the work it is delivering to each compute node, so that slower nodes are not overwhelmed, and faster nodes can operate at maximum performance.

\subsubsection{Hit-less reconfiguration of LB table}
When the offered compute nodes are altered, the load balancer needs to adjust the selection mechanism for current, and future atomic groupings of packets that need to land on a single compute node.  This occurs when nodes are added or deleted from a running cluster, as well as when the weights are changed based on dynamic feedback.  These adjustments need to be made without dropping, or mis-directing a single packet.

\subsubsection{Unidirectional UDP streaming}
Unidirectional UDP streaming, with no feedback or back-pressure meets the requirements of this use case.  At these very high data rates, with very large latency across the wide area network, it is difficult to maintain transmit buffers at the DAQs for re-transmission, or congestion management.  If there is a loss of average bandwidth in the network, it is impossible to maintain the real time transport requirement.

\subsection{Multiple Virtual Load Balancing Contexts}
Multiple independent experiments may require load balancing services at any given time.  Each experiment requires an independent allocation of Compute Nodes running distinct analysis on the data being produced.  The load balancer supports multiple IPv4 and IPv6 addresses, with each destination address mapping to one of four independent instances of all of the load balancing context.  This is further described in the P4 table structure in fig \ref{fig:LB P4 table structure}.

\section{LB Protocol Description}
The load balancer design introduces a LB Protocol header after the UDP header in a standard IP/UDP frame.  This header, as well as the IP/UDP header are terminated at the load balancer.  The payload is extracted and transmitted by the load balancer using a new IP/UDP header that defines the location of the selected compute node (CN).

\begin{figure}
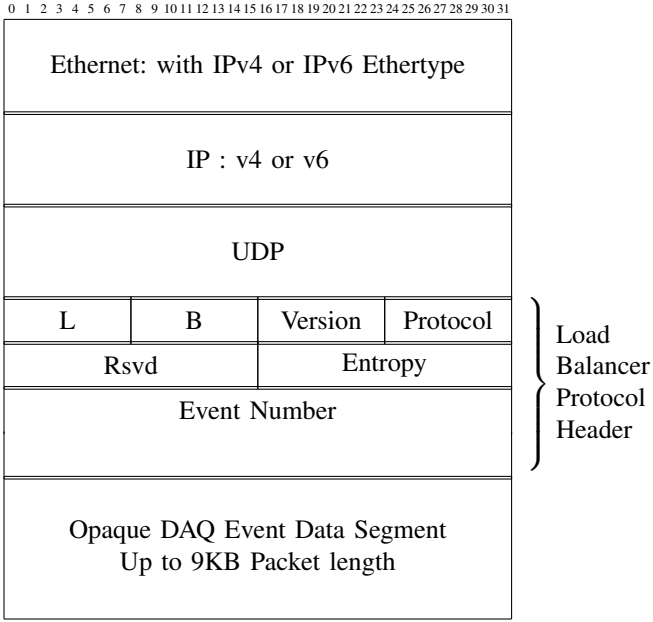

    \centering
\begin{bytefield}[endianness=little,bitwidth=.6em]{32}
    \bitheader{0-31} \\
    \wordbox{2}{ Ethernet: with IPv4 or IPv6 Ethertype } \\
    \wordbox{2}{ IP  : v4 or v6 } \\   
    \wordbox{2}{ UDP  } \\

    \begin{rightwordgroup}{Load \\ Balancer \\ Protocol \\ Header}
        \bitbox[tblr]{8}{L} & \bitbox[tblr]{8}{B} & \bitbox[tblr]{8}{Version} & \bitbox[tblr]{8}{Protocol} \\
        \bitbox[tblr]{16}{Rsvd} & \bitbox[tblr]{16}{Entropy} \\
        \bitbox[tlr]{32}{Event Number} \\
        \bitbox[blr]{32}{} 
    \end{rightwordgroup} \\
    
    \wordbox[tlrb]{3}{Opaque DAQ Event Data Segment 
                   \\ Up to 9KB Packet length } 
\end{bytefield}

    \caption{Load Balancer Packet Format from DAQ to Load Balancer }
    \label{fig:lb_packet_format}
\end{figure}

The LB Protocol header, is described in fig \ref{fig:lb_packet_format}.  The magic bytes LB, version, protocol and rsvd, provide for validation and future expansion of the LB header.  The two remaining fields "Event Number" and "Entropy" provide all the information needed for the load balancer to operate.

\subsection{LB Event Number field} The Event Number field is a monotonically increasing 64 bit value during the course of an experiment.  The monotonic nature of this field allows the event sequence space to be divided into epochs with arbitrary boundaries (e.g. 1900 $<$ Event Number $<$ 1930).  All packets for a given event must carry a common Event Number field value which requires some level of synchronisation between data sources.  This common Event Number field value anchors the entire event within a particular epoch.  Each Event Number value represents an "atomic" grouping of packets which must be delivered to a single compute node.  This "atomic" delivery can be achieved statelessly as all events within an epoch are load balanced using a stable, unchanging mapping onto the CNs which are active during that epoch.

Due to the way that the Event Number is used when choosing among CNs, it is important to choose an Event Number which has the property that the value in the 9 lsbs of the Event Number are all equally likely to occur. Without this property, the load balancing will not be statistically even across CNs.

A common method to assign an Event Number is to use the high resolution timestamp from the DAQ trigger as the Event Number value.  In synchronized DAQs, all the individual A/D converters would produce the same Event Number value, and a single compute node would be able to assimilate the data into a single event.

The load balance pipeline is built using the P4 language compiler \cite{b1} for Verilog.  In P4, range based table lookup is not defined.  Instead the ranges are programmed using all of the prefix matches to build up each range.

\subsection{LB Entropy Field} After selecting a specific CN using the Event Number field, the Entropy field is used to select from a number of different UDP ports that the load balancer will transmit to.  The port ranges can be specified independently for each compute node.  They are provided as a base UDP port address that the CN will listen on, as well as a contiguous range expressed as a power of 2.  For example (base = 1000 , range = $2^{N}$).  The load balancer will spread traffic to UDP ports at base+0 , base+1 .. base+$2^{N}$-1.  Due to limitations of the P4 language the ranges need to be powers of 2.

This mechanism enables a form of RSS (Receive Side Scaling).  It allows the CN software to run independent UDP receivers on different cpu cores, avoiding the bottleneck of a single core packet reassembly process.

The complete transformation of the packet received from the DAQ systems, into the packets transmitted to the compute nodes, is summarized in fig \ref{fig:lb_packet_rewrite}.

\subsection{Payload Segmentation and Reassembly} The Event Data Bundles constructed by the DAQs are typically much larger than the 9KB maximum network packet size.  Since the load balancer processes each Ethernet frame in isolation, and that processing depends on the presence of the Load Balancer Protocol header in every packet, IP fragmentation cannot be used to split the Event Data Bundle for transmission.

A dedicated, application layer segmentation and reassembly protocol is required.  This protocol runs between the DAQ and the compute node. The load balancer does not participate in this protocol.  Even though this segmentation and reassembly protocol is entirely opaque to the load balancer, care must be taken to properly assign related segments the same Event Number and Entropy values to ensure that the load balancer forwards them all to a single CN and also to a single UDP receiver on that CN.

A DAQ Event Data Bundle is assigned an Event Number.  Independently for each DAQ Event Data Bundle, a DAQ segments its Data Bundle into a set of Data Segments such that each Data Segment along with all headers fits within the 9KB maximum network packet size.  Each of the Data Segments generated from a single Event Data Bundle carries the same assigned Event Number in its Load Balancer Protocol header.

Further, an Entropy field is determined (randomly or sequentially) to be used in all of the Segments generated from a single Bundle.  This additionally ensures that all Segments for this Bundle will be delivered to the same CN UDP receiver to be reassembled.


\begin{figure}
    \centering
    \small
\resizebox{\columnwidth}{!}{%
\begin{tblr}{|l|l|}
        \hline
        Received from DAQ & Sent to Compute \\
        \hline 
        \hline
        Ethernet & Ethernet \\
        \hspace{1pt} ETH DA = LB MAC DA & \hspace{1pt} ETH DA = Next HOP MAC DA     \\
        \hspace{1pt} ETH SA = Any       & \hspace{1pt} ETH SA = LB MAC Addr         \\
        \hspace{1pt} ETH type = IPv4 or IPv6 & \hspace{1pt} ETH type = IPv4 or IPv6 \\
        IP & IP \\
        \hspace{1pt} IP SRC = Any        & \hspace{1pt} IP SRC = LB Addr            \\
        \hspace{1pt} IP DST = LB Addr    & \hspace{1pt} IP DST = Compute Node Addr  \\
        \hspace{1pt} IP PROT = UDP       & \hspace{1pt} IP PROT = UDP               \\
        UDP & UDP \\
        \hspace{1pt} DST PORT = LB SVC PORT (19522) & \hspace{1pt} DST PORT = CN Rx Port (1..N)  \\
        \hspace{1pt} SRC PORT = Set by DAQ          & \hspace{1pt} SRC PORT = Set by DAQ \\
        \hline[dashed]
        LB Protocol Header & \SetCell[r=2]{} Payload \\
        \cline[dashed]{1}
        Payload \\
        \hline 
\end{tblr}
}
    \caption{Load Balancer Packet Rewrite Description }
    \label{fig:lb_packet_rewrite}
\end{figure}

\section{LB P4 Implementation}
The match action table structure for a P4 implementation of the load balancing protocol is summarized in fig \ref{fig:LB P4 table structure}.

\begin{figure}
    \centering
    \small
\resizebox{\columnwidth}{!}{%
\SetTblrInner{ vlines, vspan=default}
\begin{tblr}{
             column{1}={0.37\columnwidth , font=\footnotesize},
             column{2}={0.63\columnwidth , font=\footnotesize},
             cell{1}{1}={c=2}{c},    
             cell{2}{2}={r=5}{l},
             cell{7}{1}={c=2}{c},    
             cell{8}{2}={r=7}{l},
             cell{15}{1}={c=2}{c},   
             cell{16}{2}={r=5}{l},
             cell{21}{1}={c=2}{c},   
             cell{22}{2}={r=6}{l},   
             cell{28}{1}={c=2}{c},   
             cell{29}{2}={r=9}{l}.
            }
  \hline
  {\bf Layer 2 Input Filter} \\
  \hline
  {\bf Key:} & Rejects any packets not destined for any of the load balancer's configured unicast, multicast or
                broadcast Ethernet MAC addresses.  For accepted packets, this table also selects a preferred
                unicast MAC address to be used in any output packets generated by the received packet. \\
  \hspace{10pt}Input Port (wildcard) & \\
  \hspace{10pt}Ethernet MAC DA & \\
  \hline[dashed]
  {\bf Value:}  & \\
  \hspace{10pt}LB Unicast MAC SA & \\
  \hline
  \hline
  {\bf Layer 3 Input Filter} \\
  \hline
  {\bf Key:} & Rejects any packets not destined for any of the load balancer's configured unicast, multicast or
                broadcast IPv4 or IPv6 addresses.  For accepted packets, this table also selects a preferred 
                unicast IP address to be used in any responses generated by each received packet.  Further,
                accepted packets are assigned to the appropriate LB Instance ID for further processing. \\
  \hspace{10pt}Input Port (wildcard) & \\
  \hspace{10pt}Ethertype & \\
  \hspace{10pt}IPv4/v6 Dst or ARP TPA & \\
  \hline[dashed]
  {\bf Value:}  & \\
  \hspace{10pt}LB Unicast IP Src & \\
  \hspace{10pt}LB Instance ID & \\
  \hline
  \hline
  {\bf Calendar Epoch Assignment} \\
  \hline
  {\bf Key:} & Divides the incoming Event Number space into discrete regions called Epochs.  Each defined
                Epoch has a stable, immutable Load Balance Calendar.  Past, Current and Future Epochs are
                all active simultaneously, allowing ticks in the tail of the Current to arrive intermingled
                with the head of the Next. \\
  \hspace{10pt}LB Instance ID & \\
  \hspace{10pt}LB Event Number (lpm) & \\
  \hline[dashed]
  {\bf Value:} & \\
  \hspace{10pt}Calendar Epoch & \\
  \hline
  \hline
  {\bf Calendar to Member Map} \\
  \hline
  {\bf Key:} & Provides a stable, weighted mapping from Event Numbers onto Load Balance Members available
                during this Epoch.  The Calendar Epoch selects the transition boundary between the current
                and future LB configuration.  The 9 lsbs of the Event Number select between Compute Nodes.
                Up to 512 nodes are supported for the present use case.  The number 512 can be raised to larger
                sizes by examining more than 9 lsbs in the 64 bit Event Number field.  With fewer nodes, the 
                nodes can be repeated to fill the 512 slots.  The number of repeats determines the ratio of 
                traffic one node gets relative to others.\\
  \hspace{10pt}LB Instance ID & \\
  \hspace{10pt}Calendar Epoch & \\
  \hspace{10pt}LB Event Num \& 0x1FF & \\
  \hline[dashed]
  {\bf Value:}  & \\
  \hspace{10pt}LB Member ID & \\
  \hline
  \hline
  {\bf Member Lookup and Rewrite} \\
  \hline
  {\bf Key:} & Maps from a selected Member ID to the Ethernet, IP and UDP destination for the selected Compute Node. 
                IPv4 input packets are sent to IPv4 Compute Node addresses.  IPv6 input packets are sent to IPv6
                Compute Node addresses. \\            
  \hspace{10pt}LB Instance ID & \\
  \hspace{10pt}EtherType (IPv4/v6) & \\
  \hspace{10pt}LB Member ID & \\
  \hline[dashed]
  {\bf Value:}  & \\
  \hspace{10pt}MAC DA next hop router  & \\
  \hspace{10pt}IPv4/v6 Dst & \\
  \hspace{10pt}UDP Dst Base Port & \\
  \hspace{10pt}Entropy Bit Mask Width & \\
  \hline
\end{tblr}
}
    \caption{Load Balancer P4 Table Structure }
    \label{fig:LB P4 table structure}
\end{figure}

\subsection{Parsing Stage}
The following protocols are parsed by the Parsing stage of the P4 implementation:
\begin{itemize}
    \item ARP
    \item IPv4 / IPv6
    \item ICMP Echo (Ping) Request
    \item IPv6 ND Neighbor Solicitation
    \item UDP
    \item EJFAT UDP Load Balancer Protocol
\end{itemize}

EJFAT UDP Load Balancer Protocol (LB) packets are identified based on a UDP destination port of 19522 (0x4c42, 'LB').  The Parser validates that both the Magic and Version fields in LB packets match expected values.  A mismatch from the expected values results in the packet being discarded.

No parsing is done on any bytes beyond the LB Protocol header.

\subsection{Load Balancer Initialization}

While the Load Balancer supports more complex configurations, the following subsections depict how the tables in the P4 implementation are programmed to realise the following simple example configuration of the Load Balancer.
\begin{itemize}
    \item Single Virtual Load Balancer Instance (ID 0)
    \item Static LAG across all 100G ports
    \begin{itemize}
        \item Note the wildcard (*) in the input port matches for L2 and L3 Filter Tables
    \end{itemize}
    \item Single Unicast MAC Address used across all 100G ports
    \item All IPv4 and IPv6 addresses are mapped to LB ID 0
\end{itemize}

\subsubsection{Populate L2/L3 Input Filter Tables}
The Layer 2 and Layer 3 Input Filter tables are responsible for rejecting any received packets that are not intended for the Load Balancer.  In particular, this rejects any packets that may be flooded by the network and arrive on the 100G ports of the Load Balancer even though they are destined for another host.  In order to participate as an IPv4 and IPv6 host in a network, the Load Balancer must accept packets destined to various Unicast, Multicast and Broadcast addresses at both Layer 2 and at Layer 3.  The entries in these tables enable the primary load balancing function of the device but also autonomous participation in ARP, IPv6 Neighbor Discovery and ICMP Echo/Ping protocols which are also implemented in the P4 pipeline.

Since the Load Balancer may have multiple "identities" in the network (one for each Virtual Load Balancer Instance), the rules in each table the source address which should be used in any response packets that may be generated.  Further, the L3 table provides the Load Balancer Instance ID to set the context for all further pipeline processing.

These tables typically remain unchanged after initialization since they describe the link layer and network layer addresses of the Load Balancer.

L2 Input Filter Entries
\begin{itemize}
  \item Broadcast MAC
  \begin{itemize}
    \item Key: (*, Broadcast MAC) \\
        Val: (LB MAC SA)
  \end{itemize}
  \item Unicast MAC
  \begin{itemize}
    \item Key: (*, LB Unicast MAC) \\
        Val: (LB MAC SA)
  \end{itemize}
  \item IPv6 Solicited Node Multicast MAC
  \begin{itemize}
    \item Key: (*, 33:33:FF:pp:qq:rr) \\
        Val: (LB MAC SA) \\
        Where: ppqqrr = 24 lsbs of IPv6 Unicast IP
  \end{itemize}
\end{itemize}

L3 Input Filter Entries
\begin{itemize}
  \item Unicast IPv4
    \begin{itemize}
        \item Key: (*, 0x0800, LB IPv4 Dst) \\
            Val: (LB IPv4 Src, LB ID 0)
    \end{itemize}
  \item ARP
  \begin{itemize}
    \item Key: (*, 0x0806, LB IPv4 Dst) \\
        Val: (LB IPv4 Src, LB ID 0)
  \end{itemize}
  \item Unicast IPv6
  \begin{itemize}
    \item Key: (*, 0x86dd, LB IPv6 Dst) \\
        Val: (LB IPv6 Src, LB ID 0)
  \end{itemize}
  \item IPv6 Solicited Node Multicast IP
  \begin{itemize}
    \item Key: (*, 0x86dd, ff02::0001:FFpp:qqrr) \\
        Val: (LB IPv6 Src, LB ID 0) \\
        Where: ppqqrr = 24 lsbs of IPv6 Unicasts IP
  \end{itemize}    
\end{itemize}

Result : Any packet that is not destined to the load balancer will be discarded.  This table sets up the network L2 and L3 addresses for the load balancer.

\subsubsection{Populate Member Lookup and Rewrite Table}
Establishing a new configuration for a new Epoch starts from the end of the P4 pipeline and builds toward the start.  This ordering is important to ensure that all of the downstream tables are fully populated prior to activating a new Epoch.

First, the set of Load Balance Members (the CNs) is programmed.  Each of the members will require a unique Member ID, scoped within the Load Balancer Instance ID.  The Member IDs are allocated by the control plane.

For each LB Member allocate a free Member ID and insert the applicable entry or entries for the CN.
\begin{itemize}
  \item Key: (LB ID 0, 0x0800, LB Member ID) \\
        Val: (IPv4 Rewrite, Next-Hop MAC DA, CN IPv4 Dst, CN Base UDP Dst Port, CN Dst Port Entropy Bits)
\end{itemize}
\begin{itemize}
  \item Key: (LB ID 0, 0x86dd, LB Member ID) \\
        Val: (IPv6 Rewrite, Next-Hop MAC DA, CN IPv6 Dst, CN Base UDP Dst Port, CN Dst Port Entropy Bits)
\end{itemize}

\subsubsection{Populate Load Balance Calendar to Member Map}
Once all of the LB Member entries are programmed, configuration proceeds to the Load Balance Calendar.  This table controls the weighted distribution of Event Numbers across the set of Load Balance Members available within this Epoch.

A Load Balance Calendar is a unique number scoped within an Epoch within a particular Load Balancer Instance ID.  The Calendar Epoch is allocated by the control plane.

All (or any subset of) the Member IDs available to this Load Balancer Instance ID should be distributed into the 512 Calendar Slots available in the Calendar.  Any members can occur between 0-512 times in the calendar.  A member occurring more times in the calendar has a higher “weight” and will be more likely to be assigned an event within this Calendar Epoch.

NOTE: All 512 slots MUST have a member assigned to them or events that target the empty slot will be entirely discarded by the load balancer.

For each of the 512 Calendar slots program :
\begin{itemize}
  \item Key: (LB ID 0, Calendar Epoch, Calendar Slot)
  \item Val: (LB Member ID)
\end{itemize}

\subsubsection{Populate the Calendar Epoch Assignment Table}
Once the Load Balance Calendar for this Epoch has been fully populated, it is now ready to be connected into the Event Number space.  This connection is done by programming one or more entries into the Calendar Epoch Assignment Table which describe the boundaries of the beginning (and optionally the end) of the Epoch.  During initialization, it is sufficient to map all possible Event Numbers into the first Epoch.  This end of this Epoch can be programmed at some future time in order to activate a new Epoch.

Assign the entire Event Number space to the newly allocated Calendar Epoch.

\begin{itemize}
  \item Key: (*)
  \item Val: (Calendar Epoch)
\end{itemize}

Result: The load balancer will now assign each received packet to exactly one of the LB members based on the Event ID contained in the packet.  The mapping will remain consistent for any given Event ID within an Epoch since the Calendar and Member tables cannot change within a given Epoch.

\subsection{Changing The Load Balancer On the Fly}

This section assumes that the load balancer is in-service and as such, care must be taken to avoid service disruption during reconfiguration.  If the load balancer is out-of-service, it can be reconfigured using the initialization steps above without care for disruption.

Any Epoch that is reachable (connected) via the Calendar Epoch Assignment table MUST not be changed.  In-service reconfiguration of the load balancer is done by the following steps.
\begin{itemize}

\item Allocate the next free Calendar Epoch ID scoped by the Load Balancer Instance.
\item This new Calendar Epoch ID will be activated as the last step in this reconfiguration process.
\item Insert new entries into the Load Balance Member Table for any CNs that need to be changed in the next Epoch
\item Compute and insert an entirely new calendar into the Load Balance Calendar to Member Map Table using the next Calendar Epoch ID
\item Choose an Event ID in the (near) future which will become the boundary between the current Epoch and the new Epoch.
\item Compute a set of LPM prefix matches over the Event ID space which describe the entire range of Event IDs from the start of the current Epoch up to the start of the new Epoch.
\item Program the LPM prefix matches into the Calendar Epoch Assignment Table
\item Update the wildcard match in the Calendar Epoch Assignment Table to point to the new Epoch
\end{itemize}

Result: The new Epoch is activated and its LB Calendar and LB Members MUST NOT be changed.

After waiting an appropriate time for all events from the previous Epoch to have quiesced, perform the following cleanup steps.
\begin{itemize}
\item Delete the LPM prefix matches for the previous Epoch from the Calendar Epoch Assignment Table.  This disconnects all references for the previous Epoch to the rest of the pipeline tables.
\item Delete the LB Calendar for the previous Epoch
\item Delete any unreferenced LB Member rewrites for the previous Calendar
\end{itemize}

\section{LB Hardware Implementation and Testing}
\subsection{Hardware Architecture}

\begin{figure}
    \centering
    \resizebox{\columnwidth}{!}{%
        \includegraphics[]{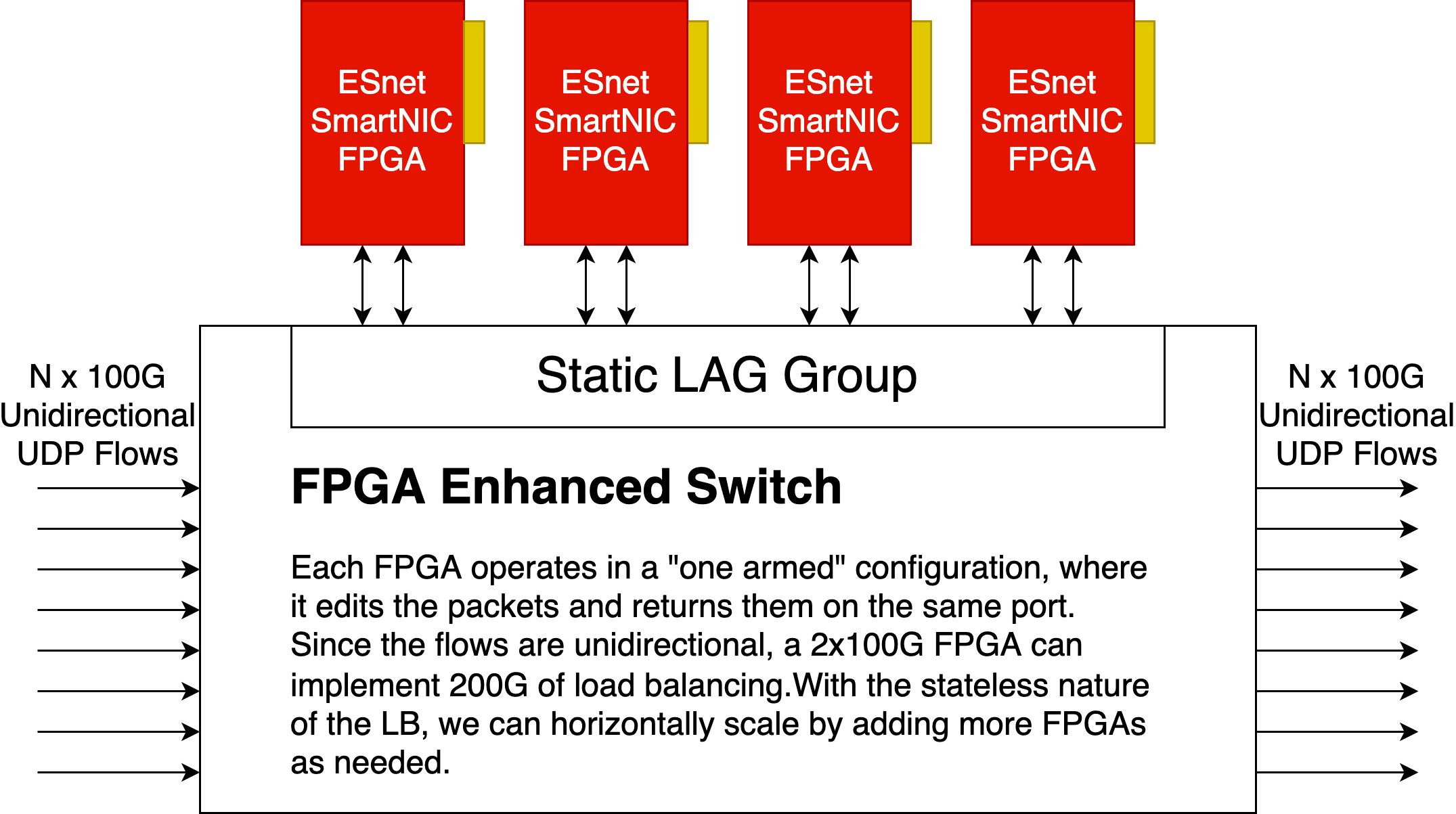}
    }
    \caption{P4 enhanced packet processing using ESnet SmartNIC built on AMD-Xilinx Alveo cards}
    \label{fig:LB_FPGA_Switch}
\end{figure}

The FPGA based load balancer is paired with a conventional Layer 2 switch to provide a horizontally scalable solution, that can grow to support any necessary bandwidth in increments of 100Gbps.  This is illustrated in fig. \ref{fig:LB_FPGA_Switch}. It is worth noting, that adding FPGAs allows us to scale the total bandwidth between data centres, and is independent of the number of compute nodes.

Using multiple Link Aggregation Groups (LAGs), the switch can be used to provide multiple FPGA groups for independent use cases.  In the future, the concept can be expanded to  run load balancing on one group, whilst adding in data compression, shaping or segmentation and reassembly with different FPGAs running different pipelines, in a cascade.

\subsection{Test environment}
\subsubsection{System level testing}

\begin{figure}
    \centering
    \includegraphics[width=\columnwidth]
       {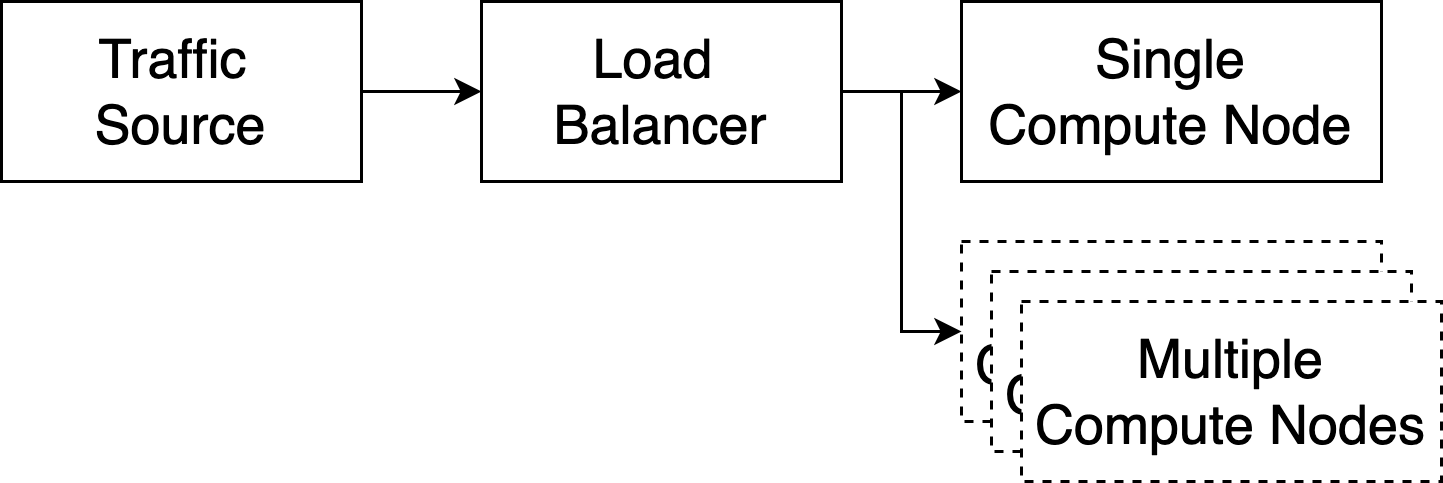}
    \caption{LB System Test Configuration : 
      \it For performance testing either a single or small number of compute nodes are used.  To analyze a large number of compute nodes, the LB tables can be configured to spread traffic over a large number.  The vast majority of the nodes can be absent, simply discarding the traffic to them.  Whilst a single CN can be used to analyze the subset of traffic destined to it.
    }
    \label{fig:LB_Sys Test}
\end{figure}

For system level testing of the load balancer to ensure that packets are being forwarded correctly, as well as to verify packet loss, hit-less switching, traffic shape and other attributes, we connect the load balancer as shown in fig. \ref{fig:LB_Sys Test}.

\subsubsection{P4 Testing}  The P4 implementation of the load balancer can be verified using a software based P4 simulator supplied with the AMD-Xilinx P4 compiler.  This allows us to generate .pcap files with carefully crafted packets which serve as the input to the simulator.  An output .pcap file is produced by the simulator, which can be examined via. wireshark, to confirm that the P4 code is correctly editing and discarding each packet.

\subsubsection{Stand alone FPGA line rate testing}  
\begin{figure*}
    \centering
    \includegraphics[width=\textwidth]
       {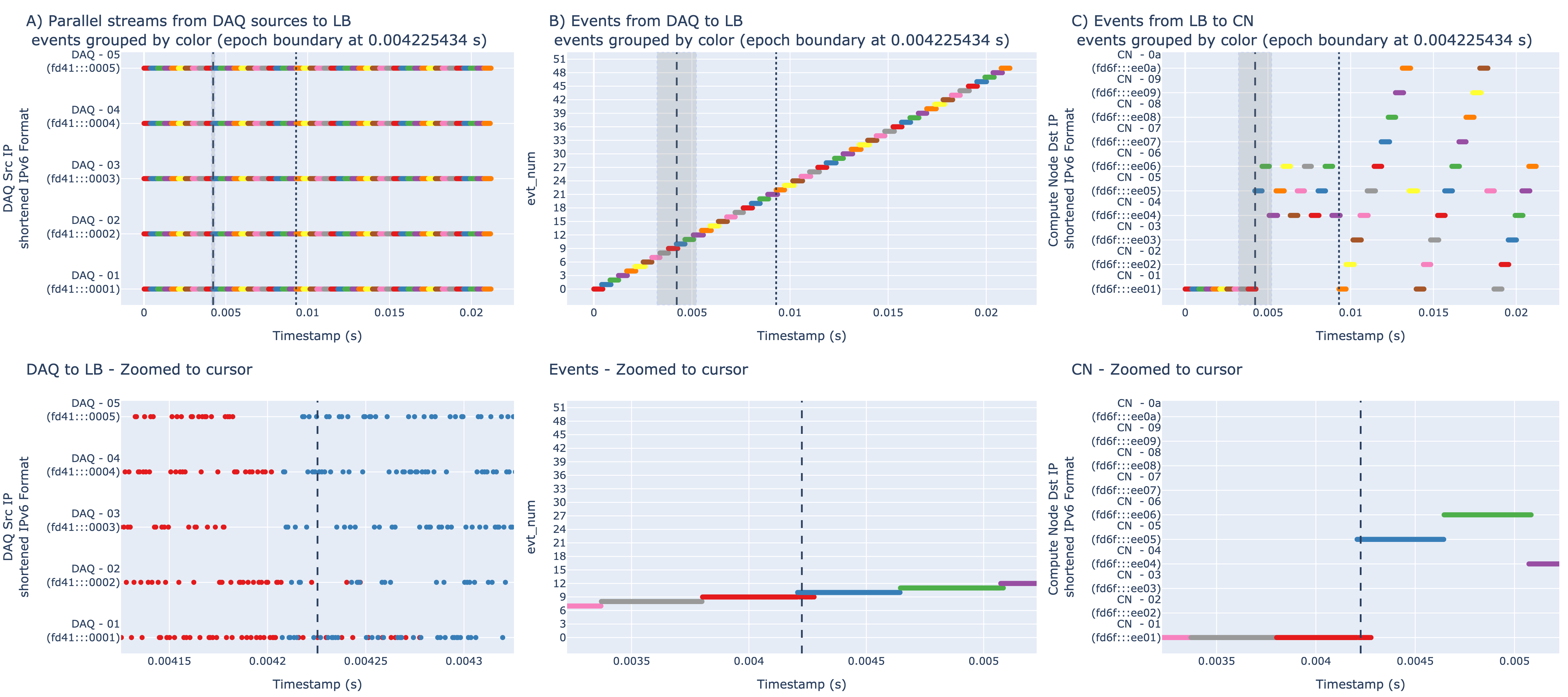}
    \caption{LB System Test Results : \it DAQ traffic generated using a DPDK DAQ emulator, streaming into a Xilinx U280 FPGA card.  Load balancing done using EJ-FAT implementation on the U280 card, and CN packet capture performed using tcpdump and hardware timestamping on a Mellanox ConnectX-5 card.  10 virtual CNs are emulated on a single 100G server.  All tests performed at a rate of 98 Gb/s.
    }
    \label{fig:LB_Test_Result}
\end{figure*}

A single FPGA with 10 virtual compute nodes was tested for FPGA capacity and implementation verification.  The test data was sourced from .pcap files on the FPGA host system, and passed directly into the load balancer via DPDK over the PCIe bus.  The .pcap files were configured to emulate 5 DAQs, as well as some network delay and reordering between the DAQ and the LB input ports.  Streaming data from the compute nodes was captured using tcpdump operating at 100Gbps.  The .pcap captures were then analyzed and plotted using the Python Pandas library.  The EJ-FAT load balancer supports both IPv4 and IPv6, for our testing we selected IPv6.  DAQs and CNs are placed in separate subnets, to demonstrate the goal of having independent name spaces for infrastructure.  Measured test data is shown in fig. \ref{fig:LB_Test_Result}.

\subsection{Real Time Test Results}The subplot in fig. \ref{fig:LB_Test_Result}a shows traffic sourced from 5 different virtual DAQs, each with a unique IP address.  The aggregate bandwidth from all the DAQs is set to 98Gbps, in order to establish a non-congesting end to end data path.  Events are sequentially numbered by the DAQ systems, and represent correlated data that needs to be sent to a single CN for reassembly.  Event boundaries are shown using distinct colours.  Below this figure we show a close up of the packet data around the cursor boundary.  Individual packets are visible, and the packet re-ordering between adjacent events can be seen.

It can be seen that each DAQ provides a variable number of total data samples per event.  This is evident by the fact that some DAQs finish before others, for a given event number.  Events are typically synchronized across DAQ systems by a hardware trigger system that is synchronized to within 1ns.

The subplot in fig. \ref{fig:LB_Test_Result}b shows the packets as seen by the load balancer after passing through a network and emulated aggregation switch.  Packet serialization and random path delays are built into the traffic generator presenting traffic directly at the load balancer FPGA input.  It can be seen that despite the fact that the DAQs transmit parallel event data, once it arrives at the load balancer event data is grouped together in a single burst, with some overlap at the edges of an event, due to network packet reordering.  Event numbers are monotonically increasing, which is a key requirement for the epoch forwarding table behaviour.

The subplot in fig. \ref{fig:LB_Test_Result}c shows the packets as seen by the compute nodes, at the output of the load balancer.  For this test we introduced 3 epochs, where the number of assigned compute nodes is changed during the experimental run.  We begin with just 1 CN assigned.  In the next epoch we increase the number of compute nodes to 3, and we add new compute nodes CN-4, CN-5 and CN-6, and we remove CN-0.  For the final epoch we enable all 10 CNs, and allow the load balancer to operate for several full event cycles, showing the fair distribution of sequential events to all compute nodes.  As well we increase the weight for CN-5 so that it is assigned more work than the rest.

For correct behaviour, in this last plot, we are looking to ensure that events are not split across epoch boundaries into two different compute nodes.  As well we conducted a full accounting between input and output .pcap files to ensure that no packets are lost during the test, particularly during the epoch switching events.  This was confirmed using 9000 byte UDP packets and at a line rate of 98Gbps.  The CN's were equipped with Mellanox ConnectX-5 100G NICs.  The tcpdump application with hardware timestamping was used to measure the traffic.  The DAQ traffic emulation was done using DPDK driving a U280 FPGA card.

\section{Comparison with Prior Solutions}

Barefoot networks has developed an FPGA based P4 load balancer \cite{b4}, that is subtended from a network switch.  It is similar to the EJ-FAT solution, in that it uses a high bandwidth switch, to subtend a number of FPGA devices. The EJ-FAT approach is different for a number of reasons.  In the Barefoot design, the switch is P4 programmable and implements a pre-cached sorting function for active flows.  EJ-FAT on the other hand works with all switches, and relies on industry standard, stateless Link Aggregation for the spreading function.  Having no special requirements on the switch is very desirable in production environments.  Another key difference is the absence of a specific load balancing header in the Barefoot solution.  It instead makes a multi level decision using L3 packet headers and ACLs.  For real time workflows from DAQs, we have described the benefits of a new LB header, which provides a location/id separation between data centres.  As well the new load balance header provides application driven data grouping to specific CNs.  The Barefoot load balancer needs table sizes proportional to the number of flows, whereas the EJ-FAT load balancer needs table sizes proportional to the number of compute nodes.

The TIARA load balancer described in \cite{b5}, shares some attributes with both the EJ-FAT approach and the Barefoot approach.  They use ECMP to perform a stateless load balancing into an array of FPGAs.  ECMP is a L3 version of L2 LAG.  Like other server load balancing (SLB) designs their goal is to perform load balancing using just L3 headers.  They compare other designs built on programmable switches, with those that are built on FPGAs, and recognize the scalable design that comes from stateless load balancing into as many FPGAs as needed, whilst leveraging large High Bandwidth Memory (HBM) for maintaining flow state in the FPGAs.  In EJ-FAT, due to the use of an application aware header, and a mapping that only scales with the number of CNs, the EJ-FAT implementation is done using a very small number of FPGA block RAM, with no need for HBM.  This significantly drives down the cost and power of the FPGAs as they scale out in large numbers.

Using the two examples above, we have demonstrated the similarities and differences between Server Load Balancing (SLB) and real-time load balancing for large scientific instruments.  EJ-FAT is required to perform real-time load balancing, and not server load balancing.  Thus explaining some of the key complexities that Barefoot, and TIARA undertake relative to EJ-FAT.  One of the largest instruments is the Large Hadron Collider at CERN.  In \cite{b6} we see a framing of the problem that EJ-FAT is designed to tackle.  In most accelerators, there is a design principle towards doing data reduction and event filtering as early as possible in the data flow.  This has led to DAQ systems tightly coupled with large banks of CNs at the accelerator site.  Within DOE there are a large number of comparable instruments. This leads to inefficient placement of CNs, which are each individually sized for each accelerator, but used only when the accelerator is active.  EJ-FAT is designed to move the front end processing across the WAN to a shared pool of resources.

EJ-FAT is guided by the accelerator requirements at JLab.  In \cite{b7} the current DAQ and Compute Node architecture is described.  Due to the proximity of the DAQs and CNs JLab has built and operated a TCP based link between the two.  In this scenario, the data rates were based on 10Gbps ethernet links, and round trip times were at $\mu$s time scales.  For the next generation systems data rates will scale to 100Gbps and 400Gbps, and it will be transported at ms latencies across the WAN.  This makes the continued use of TCP with re-transmissions, very difficult to support and scale, with one round trip time worth of buffering required at the sources.  Continuing the use of TCP whilst decoupling the sources (DAQs) and sinks (CNs) would also imply TCP termination and origination at the load balancer, which eliminates the possibility of a stateless load balancer, and introduces significant complexity in terms of maintaining flow state for all flows through the load balancer.

\section{Conclusion and Future Work}

\begin{figure}
    \centering
    \includegraphics[width=\columnwidth]{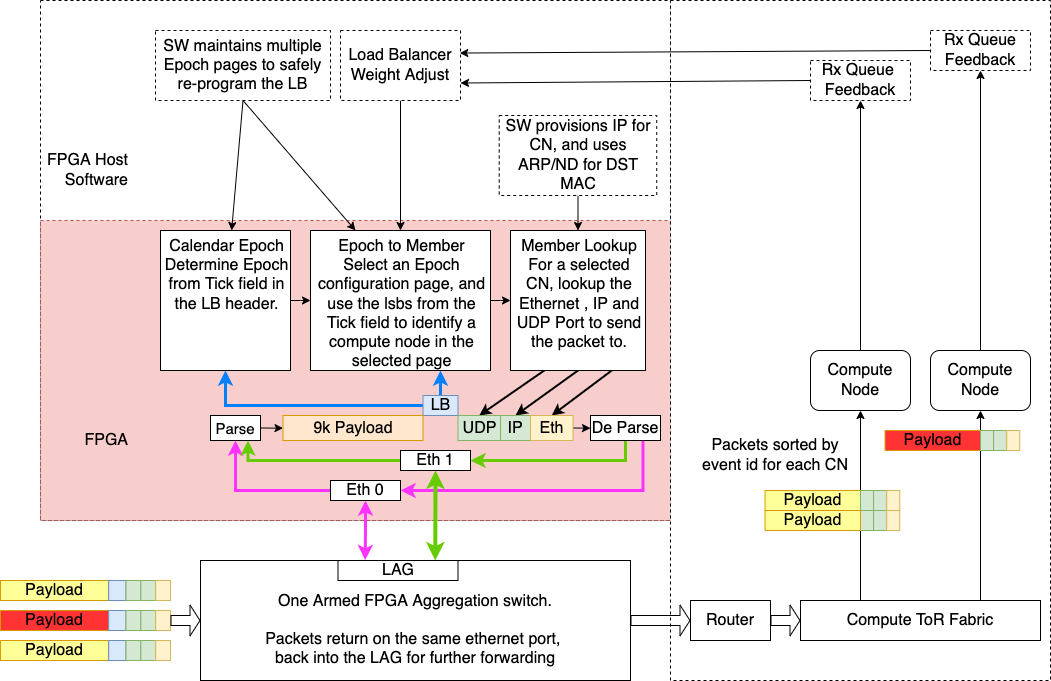}
   \caption{End to End Packet Forwarding between the Load Balancer and Compute Nodes}
   \label{fig:PacketLBtoCN}
 \end{figure}

In the prior sections of this paper, we described each of the pieces of hardware and packet processing that make up the UDP load balancer.  In fig. \ref{fig:PacketLBtoCN}, we show the integrated view of packet flow from the Wide Area Network (WAN) through the load balancer into compute nodes.

Through validation testing in a hardware testbed, we have successfully streamed simulated DAQ data at 100Gbps through the load balancer and into a simulated compute node cluster. Through careful analysis of the packet stream seen by the CNs, we have confirmed fully coherent distribution across epoch boundaries in the presence of randomly delayed packets.  As well we have tested all of the load balancer packet editing functions using a P4 behavioral simulator that takes .pcap files as input, and produces .pcap files as output for inspection with wireshark.

Future capacity testing, including multiple FPGA ports, and a LAG configuration of the switch will be conducted.  Further testing on ESnet's WAN will be performed to ensure that our assumptions of packet loss over the WAN will be met.  EJ-FAT is able to create multiple virtual LBs on a single FPGA.  This has been tested in P4 simulation, and will be tested in hardware to demonstrate more than one experiment stream from different labs can be successfully load balanced into a shared compute cluster.


%

\section*{Acknowledgment}
The EJ-FAT project makes use of the Open NIC Shell (ONS) \cite{b2} produced by AMD-Xilinx Corporation for use with the Alveo U280 FPGA cards.  
We would like to thank Gordon Brebner and Chris Neely for their support and development of ONS.

We also make use of the ESnet SmartNIC project \cite{b3}, which is an open source project derived from the ESnet6 High Touch project.  The ESnet SmartNIC simplifies the load balancer implementation to a few hundred lines of P4 code, whilst inheriting all the verilog code from the High Touch project.  Without this code reuse, it would not have been possible to implement the EJ-FAT load balancer, without a substantial FPGA development team.

\ifCLASSOPTIONcaptionsoff
  \newpage
\fi



%

\newpage
%

\begin{IEEEbiographynophoto}{Stacey Sheldon}
Stacey Sheldon is presently an affiliate with ESnet, the networking provider for the US Department of Energy.  Stacey has more than 25 years of experience in the networking industry.  Stacey was the chief architect for Corsa Networks, where he designed the industry's first fully programmable OpenFlow switch.  He holds several patents on programmable switches, as well as network security appliances.  Prior to Corsa he developed System and OS Designs for Ciena's flagship 5400 converged packet optical chassis.  Stacey worked at Catena networks, where he developed embedded software for various ADSL products.  He started his career at Bell Northern Research (Nortel), developing management software for the industry leading OC-192 Transport node.  Stacey's research interests include real time embedded software, networking protocols, and FPGA accelerated algorithms.
\end{IEEEbiographynophoto}

\begin{IEEEbiographynophoto}{Yatish Kumar}
Yatish is presently an affiliate with ESnet, the networking provider for the US Department of Energy. Yatish has more than 30 years of networking industry experience. He has been involved in a number of successful startups. At Corsa Networks a leading provider of SDN switches he was the founder and CTO. Prior to that at Catena Networks, which was acquired by Ciena in 2004, he led the design team responsible for developing the industry’s lowest power ADSL and POTS codecs which allowed Catena to become the leader in ADSL retrofits in all major RBOC accounts. Prior to Catena, Yatish started his career at Nortel where he contributed to and managed the development of a number of mixed signal semiconductor projects including designs for ADSL, POTS, CDMA, Cable Modems and handsets. He holds patents in DSP architectures, and data compression and has authored papers on high level synthesis, and embedded processor design as well as contributing to the development of ITU 992.1, ANSI T1.413 and Telcordia GR909 standards. But all this pales compared to the adventure of networking in support of large science at DOE.  Yatish's research interests include real time DSP processing, processor architectures, as well as FPGA accelerated network packet processing.
\end{IEEEbiographynophoto}



\begin{IEEEbiographynophoto}{Michael Goodrich, PhD}
Dr. Goodrich has spent his career in scientific computing, simulation, and data science.  
(He has a PhD in Computational Modeling and Simulation and his research interests are in the area of computational Bayesian inference and machine learning.)
He is currently a Streaming Data Scientist in the Experimental Physics Scientific Computing Infrastructure Group, Computational Sciences and Technology, Jefferson National Accelerator Facility.

\end{IEEEbiographynophoto}

\begin{IEEEbiographynophoto}{Graham Heyes, PhD}
Dr Graham Heyes, has been Head of the Scientific Computing Department in the Computational Sciences and Technologies Division of the Thomas Jefferson National Accelerator Facility since March 2020. From 1994 to 2020 he was Group Lead for Data Acquisition Support in the Experimental Nuclear Physics Division as well as Computer Center Director from 2003 to 2006. He was the lead developer of the CODA data acquisition framework and overall computing coordinator for Experimental Nuclear Physics. He now directs technology evaluation and system architecture design of the laboratory scientific computing infrastructure and is co-chair of the Electron Ion Collider computing coordination committee.
\end{IEEEbiographynophoto}




\end{document}